\documentclass{iaus}
\usepackage{graphics}

  \checkfont{eurm10}
  \iffontfound
    \IfFileExists{upmath.sty}
      {\typeout{^^JFound AMS Euler Roman fonts on the system,
                   using the 'upmath' package.^^J}%
       \usepackage{upmath}}
      {\typeout{^^JFound AMS Euler Roman fonts on the system, but you
                   dont seem to have the}%
       \typeout{'upmath' package installed. iaus.cls can take advantage
                 of these fonts,^^Jif you use 'upmath' package.^^J}%
      }
  \else
  \fi

  \checkfont{msam10}
  \iffontfound
    \IfFileExists{amssymb.sty}
      {\typeout{^^JFound AMS Symbol fonts on the system, using the
                'amssymb' package.^^J}%
       \usepackage{amssymb}%

      }{}
  \fi

  \IfFileExists{amsbsy.sty}
    {\typeout{^^JFound the 'amsbsy' package on the system, using it.^^J}%
     \usepackage{amsbsy}}
    {}

\newsavebox{\astrutbox}
\sbox{\astrutbox}{\rule[-5pt]{0pt}{20pt}}

\title[The Interplay among Black Holes, Stars and ISM in Galactic 
       Nuclei]{Super massive black holes in spiral galaxies: HST/STIS 
observations for 3 new objects}

\author[L. Coccato {\it et al.\/}]%
{L. Coccato$^1$, 
E. Dalla Bont\`a$^1$,  
M. Sarzi$^2$, 
A. Pizzella$^1$, 
E. M. Corsini$^1$, 
\and F. Bertola$^1$} 

\affiliation{$^1$Dipartimento di Astronomia, Universit\`a di Padova,
Padova, Italy\\[\affilskip]
$^2$University of Oxford, UK}

\pubyear{2004}
\volume{222}
\pagerange{1--8}
\date{?? and in revised form ??}
\jname{The Interplay among Black Holes, Stars and ISM \\in Galactic Nuclei}
\editors{Th. Storchi Bergmann, L.C. Ho \& H.R. Schmitt, eds.}
\def\farcs{.\hspace{-2pt}''}
\begin{document}

\maketitle

\begin{abstract}
We present long-slit HST/STIS measurements of the ionized-gas
kinematics in the nucleus of three disk galaxies, namely NGC 2179, NGC
4343, NGC 4435. The sample galaxies have been selected on the basis of
their ground-based spectroscopy, for displaying a strong central
velocity gradient for the ionized gas, which is consistent with the
presence of a circum nuclear keplerian disk (CNKD, \cite{bert1998}; \cite{funes2002}) rotating around a super massive black hole
(SMBH). For each target galaxy we obtained the H$\alpha$ and [NII]
6583 \AA\ kinematics along the major axis and two 0\farcs25 parallel
offset positions. Out of three objects only NGC 4435 turned out to
have a disk of ionized gas in regular motion and a regular dust-lane
morphology. Preliminary modeling indicates a SMBH mass ($M_\bullet$)
one order of magnitude lower than the one expected from the
$M_\bullet-\sigma_c$ relation for galaxies (\cite{ferr2000};
\cite{geb2000}).
\end{abstract}

As demonstrated in Bertola et al. (1998), it is possible to detect
from ground-based observations the presence of a CNKD around a SMBH.
Its identification is possible from the study of the position velocity
(PV) diagram as done by Funes et al. (2002). From this work and
literature data we select three galaxies, namely NGC 2179, NGC
4343 and NGC 4435, which have a PV diagram
consistent with the presence of a CNKD and a central stellar
velocity dispersions ($\sigma_c$) which falls in a poorly sampled
region of the $M_\bullet - \sigma_c$ relation.

For each target galaxy we obtained with HST/STIS the H$\alpha$ and
[NII] 6583 \AA\ kinematics along the major axis and two 0\farcs25
parallel offset positions on either sides of the nucleus. The scale
was 0\farcs05 pixel$^{-1}$, the instrumental dispersion was 0.55 \AA\
pixel$^{-1}$ ($\simeq$ 25 km s$^{-1}$) and the slit width was
0\farcs2. In Fig. 1 we report the dust-lane morphology obtained by the
unsharp-masked HST images and the STIS major-axis spectra of the three
objects.

For NGC 2179 and NGC 4343 the spectrum reveals a disturbed gas
kinematics. These two objects show an irregular dust-lane morphology,
too. On the contrary, NGC 4435, the only one with a regular
kinematics, shows also a regularity of the dust-lane morphology. This
galaxy is a good candidate to model the velocity field for the
determination of $M_\bullet$. In the model we consider a constant
velocity dispersion field, an exponential profile for the flux of the
emission lines and we build the velocity field considering the
contribution of the stellar potential (measured from the
surface-brightness profile using a constant mass-to-light ratio and
spherical symmetry) and the Keplerian potential of a SMBH. Effects of
the STIS PSF, of the slit width and the bleeding of charge between
adjacent pixels in the CCD are also taken into account.

\smallskip
 We summarize our results as follows.

\begin{itemize}
\item Ground-based observations evidence that about 20\% of galaxies
show a PV diagram which is consistent with the presence of a CNKD
(Funes et al. 2002). This criterion should be combined with the
presence of a regular dust-lane morphology, according to the results
later found by Ho et al. (2002). Indeed in our sample NGC 4435 is the
only galaxy that shows a regular dust-lane morphology as well as a
regular rotation curve (Fig. 1, left panel). Combining these two
criteria we maximize the chances to select successful STIS target
galaxies for gas-dynamical measurements of SMBH masses. On the basis
of this experience we are tempted to conclude that only less than the
10\% of galaxies shows a ground-based PV diagram which is consistent
with the presence of a CNKD and has also a regular rotation curve,
useful for the measurements of the SMBH mass in a HST/STIS
spectroscopic follow-up.

\item The modeling work is still in progress (Fig.1, right panel),
but preliminary results give a $M_\bullet$ about one order of
magnitude lower than the $M_\bullet$ predicted by the
$M_\bullet-\sigma_c$ relation.  We are going to include in our model
also the available HST/STIS minor axis kinematics, and we are going to
add an exponential velocity dispersion profile and evaluate the
asymmetric drift correction, as done by Barth et al. (2001).


\end{itemize}

\begin{figure}
 \scalebox{0.33}{\includegraphics{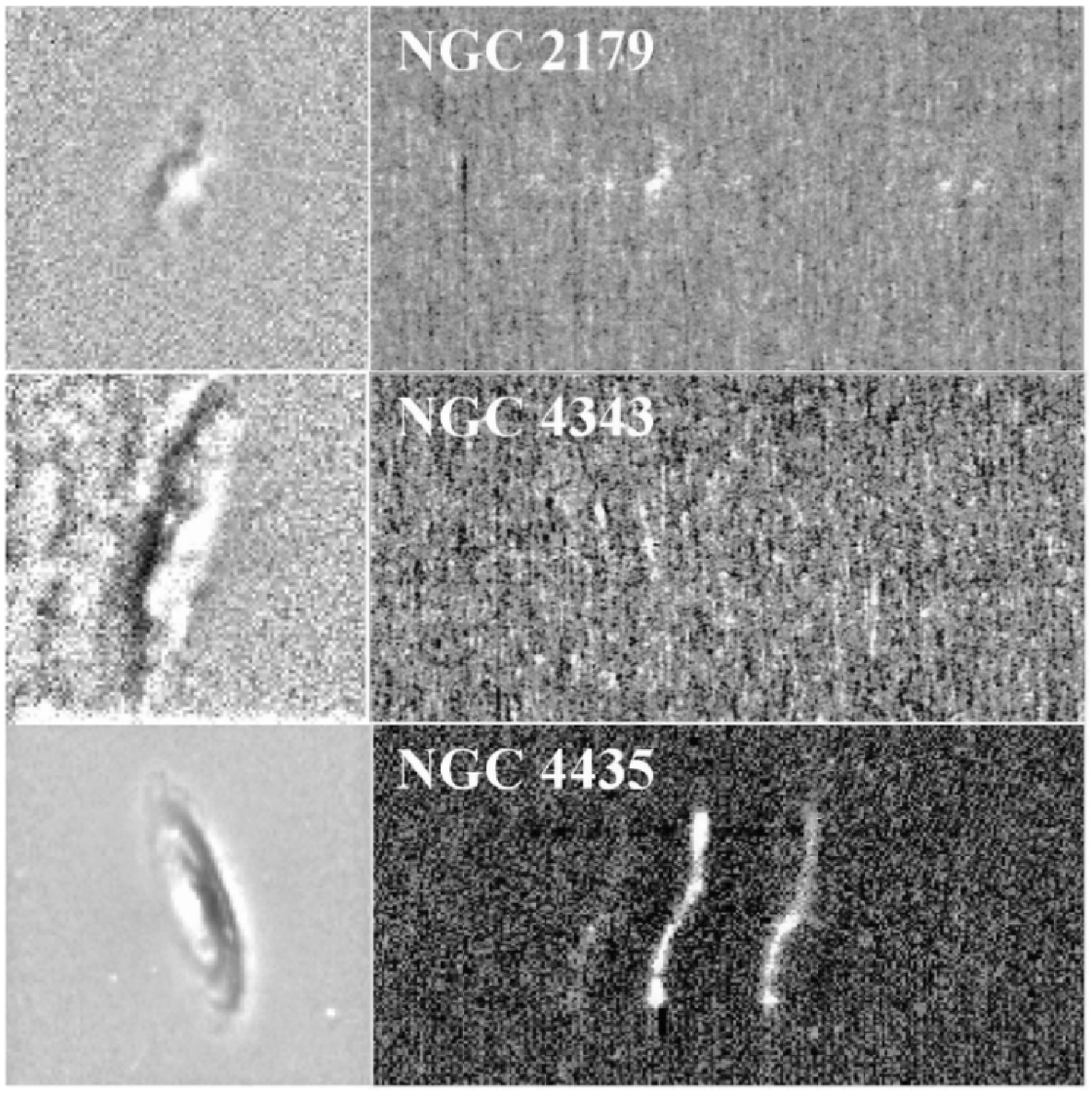}}
 \scalebox{0.33}{\includegraphics{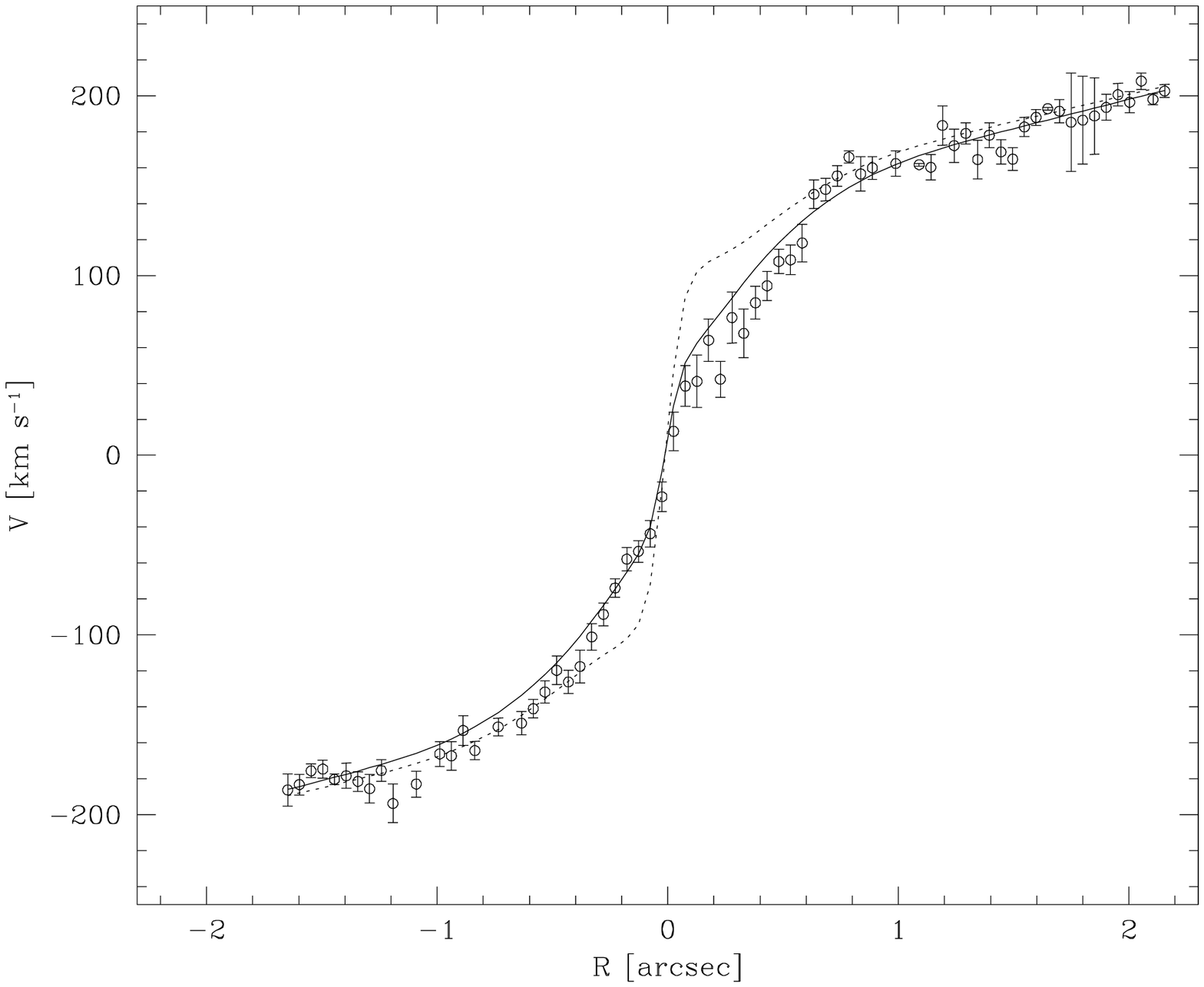}}

\caption{{\it Left panel}. Dust-lanes morphology obtained by the
unsharped-masked HST images and STIS major-axis spectra; H$\alpha$ and
[NII] emission lines are clearly visible in NGC 4435. {\it Right
panel}. Circles: observed rotation curve along the major axis of NGC
4435.  Solid line: our model; dashed line: gas rotation curve assuming
the $M_{\bullet}$ predicted by the $M_{\bullet}-\sigma_c$ relation
($M_\bullet \simeq 10^7 M_\odot$, $\sigma_c$ =174 km s$^{-1}$ from
\cite{bern2002}.)  }
\end{figure}

\end{document}